# Measuring the Earth with *Traceroute*

By
Snowflake Kicovic, Loren Webb and Michael Crescimanno

Department of Physics and Astronomy, Youngstown State University
Youngstown, Ohio, 44555

**ABSTRACT:** The *traceroute* utility on any computer connected to the Internet can be used to record the roundtrip time for small Internet packets between major Internet traffic hubs. Some of the routes include transmission over transoceanic fiber optic cable. We report on *traceroute*'s use by students to quickly and easily estimate the size of the earth. This is an inexpensive and quick way to involve introductory physics students in a hands–on use of scientific notation and to teach them about systematics in data.

**INTRODUCTION:** How big is the Earth? Ask that question in the first few lectures in an introductory general physics class and you are bound to get many answers. Invariably, some student will know 'the answer', or look it up in the course book and this then begs the question, "How do you know that's right?" Many students are surprised to learn that they can estimate the size of the earth for themselves, using only an internet enabled computer, a globe and a piece of string[1,2].

*Traceroute*, a standard utility on virtually all TCP/IP–enabled (that is, networked) operating systems, was originally developed to troubleshoot networks. This program sends out a sequence of IP packets to and from nodes (*i.e. Computers or network switches)* along the route from your computer to the designated machine. On windows it can be evoked from the MSDOS shell by typing *tracert ipname*, where *ipname* is the IP address or DNS (Domain Name Service)–resolvable name of the destination machine. Here's an example of a *traceroute* from Youngstown State University (in Ohio) to a node at the University of Hawaii.

**tracert www.hawaii.edu**
```
traceroute to 128.171.94.101 (128.171.94.101),     30 hops max, 38 byte packets
 1  ROUTER2.WB.YSU.EDU (150.134.220.1)          0.727 ms  0.561 ms  0.945 ms
 2  ROUTER1.YSU.EDU (192.55.234.11)             3.454 ms  2.489 ms  2.285 ms
 3  yqp1-atm2.youngstown.oar.net (199.18.10.37) 2.347 ms  2.603 ms  4.113 ms
 4  chi3-atm1-0.chicago.oar.net (199.18.202.173) 18.974 ms  18.829 ms  20.199 ms
 5  chicagobr.att-disc.net (206.220.243.28)     22.440 ms  23.729 ms  21.045 ms
 6  seattlebr-aip.att-disc.net (135.206.243.11) 89.214 ms  88.095 ms  89.363 ms
 7  140.32.130.186 (140.32.130.186)             137.679 ms  138.470 ms  136.947 ms
 8  harry-atm-juniper.uhnet.net (128.171.64.230) 137.975 ms  141.281 ms  138.160 ms
 9  * *
```

The student has typed only the first line (here in bold). In this case, *traceroute* is invoked using an internet name. The output of each line indicates the **round–trip** time for each independent packet to and from a node on the way to Hawaii. It is not too difficult to see that these times imply that the route chosen in this case is through Chicago and then to Seattle and then, in one big time jump, to Hawaii. Thus, **half** the difference of the average round–trip times to and from Seattle and to and from Hawaii is the additional one–way time it takes a packet to get from Seattle to Hawaii. In the above example that is about 24.4 milliseconds (ms).

But how do Internet signals get from Seattle to Hawaii? They traverse an oceanic fiber optic cable, which is buried in the mud at the ocean's floor and lies roughly along a great circle between Seattle and Hawaii. Fiber optic is made of glass, and the speed of light in the glass cable is about 2/3 of that in vacuum. This fact can be gleaned either from numerous web sites of optic cable manufacturers (which are easy to find!)[8], or through a discussion of refraction and measurement of the index of refraction in a glass sample [9] (a nice touch, but probably not what you want to do the first day of classes, for which this exersize is designed!). The speed of light moving through a fiber optic cable is basically the same as speed of propagation of an electrical signal through a computer network ('category−5 ethernet') cable. Of course, this is not an accident, but space here precludes that discussion. For the earnest student(s), we note only that the speed of propagation of a signal in a network cable can be rather directly and simply measured in a laboratory experiment using two laptops and a few network cables of different (but modest) lengths [10].

Returning to our Seattle−Hawaii transmission, assuming that most of the 24.4 ms delay is propagation in the cable, and using the relation $d=vt$ = (2/3 x 3.0 x $10^8$ m/s ) x (24.4 x $10^{-3}$ s)  = 4800 Km is an estimate of the  cable distance [11].

Assuming that your classroom globe of the earth accurately reproduces the scaled distance between points, and that the cable is laid approximately along a great circle (since that would be the shortest and thus cheapest way to lay cable) students can use ratio and proportion to convert the above measurement of the distance between Seattle and Hawaii to an estimate of the radius of the earth . We used a 15.3 cm radius globe and found a string length between Seattle, Washington and Hawaii along the surface of the globe to be about 10.4 cm long, yielding an estimate of the earth radius of 15.3*4800/10.4 = 7100 Km. Alternatively, students may use web site calculators[12] to compare the cable distance with the actual distance along the globe to again use ratio and proportion to convert their cable distance measurement to an estimate of the radius of the earth.

This is quite simple for the students to do individually or in small groups. Each group can find its own targets, and analyze a unique set of *traceroute* data to come up an estimate they contribute to a class average. Along the way they learn some basic facts about the Internet, some geography and how to read the  *traceroute* output. By far, the most difficult part of this laboratory is determining the geographic location of the nodes on either side of the transoceanic cable that you are using. Sometimes the machine names are non−descript or not given. In either case a web resource called *netgeo* is often useful[13] in translating the IP numbers to locations.

Table I below contains typical data found by some of our students between various shore points, along with the associated estimate of the earth's radius for each. As a warning, the final entry is an example between land points, where the many repeaters and non−great circle path chosen generally complicate the interpretation of the times and lead to very  poor estimates of the earth's radius.

It is noteworthy  (***) that the data displayed for New York  to Iceland  naively yields an estimate of 8600 Km for the earth radius. However, there is actually no great circle sea−route between the two sites. An obvious obstacle, Newfoundland , Canada sticks out far east into the Atlantic and precludes lying a cable along a great circle from New York to Iceland!  The Table I earth radius estimate for that datum results from  draping a string on the globe along a sea route that is entirely offshore between New York and Iceland and using its length (which for our globe was 13 cm). The apparent errors in the short hops from  the mainland  US to nearby island (in Table I, Bermuda, though Puerto Rico and other 'short hops' traceroutes are similar) may indicate that systematic delays have a proportionally larger impact on the data quality for small time differences (short routes).  Additionally, there were some sites we found that, for one reason or another (cable route unknown,  network  topology, inability to determine location of node, etc) did not work well. These include traceroutes from the USA to Fiji, Japan,  India and Italy. However, in our experience  most clear, long , ocean routes gave estimates like those of Table I. The ** for the string distance  from  Lisbon Portugal indicates that for this data the student actually used one of the web calculators alluded to earlier to compare with the cable length determined by *traceroute* ; string  and globe would give essentially the same earth size estimate.

These data for transoceanic cable routes yield estimates of the earth's radius typically some 10%−

20% too large. Clearly this indicates some systematic effect. We believe the most relevant systematic effect in this approach is that, for many reasons, the cables are not laid precisely along great circles on a perfectly spherical earth. For example, the cable is buried in mud going up and down hills at the bottom of the ocean and also around threatening ocean–bottom features.

This systematic is certain to lead to a spirited discussion about biasses in data. To help make sense of this students compiled a list of the road distances and straight–line distances between eight large Ohio cities. This may be found in Table I and a histogram of the ratio of these two distances is Figure I. Roads are expensive to build and particularly big roads between major metropolitan areas. As a result, one might expect the roads to be laid nearly along great circles (Or, on the scale of Ohio, straight lines). As all students know, that is not the case for many reasons, and it is also noteworthy that the actual road distance in this sample (which we have every reason to expect is pretty generic) is on average about 20% percent greater than the straight–line distances.

**Closing Remark s:** Besides introducing *traceroute* to students, this pedagogically straightforward class–lab can be used as a 'hands–on' exersize with scientific notation, d=vt and elementary geometry. We've had a good experience using it with students and the data quality in the lab presented here can apparently be improved somewhat with additional work[11]. As described above however, this relatively simple lab can yield atleast a crude estimate of the earth's radius, and we suspect that for students the interesting part will be reinforcing the spatial metaphor of web surfing and 'seeing' the roughness of the earth in the systematics of their data.

**Acknowledgment:** The authors thank Ron Tabak for comments on a reference and also David W. Foss for discussions. This effort was supported in part by a NASA grant NAG9–1166, a Cluster Ohio Project (Ohio Supercomputer Center) Grant and a Research Professorship 2001–2002 award from YSU. One of us (MC) is thankful to the Center for Ultracold Atoms where as a visitor this manuscript was completed. This effort was initiated and completed on equipment purchased through Research Corporation Cottrell Science Award #CC5285.

**TABLE I:** Traceroute Summaries and Associated Earth Radius Estimates

| | Departure Point | IP Address | Final Destination | IP Address | (Avg) Round Trip (ms) | Distance on Globe (cm) | Cable Distance (km) | Calculated Radius (km) |
|---|---|---|---|---|---|---|---|---|
| 1 | Los Angeles, US | 206.111.43.34 | Auckland, New Zealand | 203.97.7.69 | 139.67 | 25 | 13800 | 8450 |
| 2 | New York City, US | 146.188.179.229 | * Bermuda | 157.130.11.98 | 19.67 | 3.1 | 1950 | 9600 |
| 3 | Seattle, WA, US | 135.206.243.11 | * Hawaii | 140.32.130.186 | 48.6 | 10.1 | 4800 | 7100 |
| 4 | Los Angeles, US | 209.227.128.86 | Sydney, Australia | 209.227.148.42 | 149 | 29 | 14800 | 7800 |
| 5 | Philadelphia, PA, US | 4.24.10.181 | London, England | 195.16.175.250 | 71.33 | 15 | 7100 | 7190 |
| 6 | New York City, US | 193.251.241.217 | * Portugal | 193.251.241.133 | 76 | 14.3 | 7520 | 8040 |
| 7 | Lisbon, Portugal | 193.137.2.254 | Horta, Azores | 193.137.2.33 | 37 | --** | 3700 | 7400 |
| 8 | New York, US | 152.63.18.65 | * Iceland | 157.130.0.202 | 62.5 | 11 | 6190 | 7300*** |
| 9 | Youngstown, US | 198.18.10.37 | Chicago, US | 199.18.202.173 | 16.3 | 1.3 | 1600 | 19200 |

\* If no city listed, final destination refers to the first (coastal) city reached.

**Table II:** Comparison of Actual Highway and Straighline Distance
Between Ohio Metropolitan Areas

|    | Intial Destination | Final Destination | Staight Line Distance (km) | Quoted Distance (km) | Ratio |
|----|--------------------|-------------------|----------------------------|----------------------|-------|
| 1  | Akron     | Ashtabula  | 106  | 138 | 1.3  |
| 2  | Akron     | Cambridge  | 116  | 134 | 1.16 |
| 3  | Akron     | Cleveland  | 49.5 | 61  | 1.23 |
| 4  | Akron     | Columbus   | 173  | 229 | 1.32 |
| 5  | Akron     | Dayton     | 265  | 319 | 1.2  |
| 6  | Akron     | Toledo     | 177  | 229 | 1.29 |
| 7  | Akron     | Youngstown | 72   | 79  | 1.1  |
| 8  | Ashtabula | Cambridge  | 212  | 272 | 1.28 |
| 9  | Ashtabula | Cleveland  | 85   | 106 | 1.25 |
| 10 | Ashtabula | Columbus   | 277  | 327 | 1.18 |
| 11 | Ashtabula | Dayton     | 363  | 474 | 1.31 |
| 12 | Ashtabula | Toledo     | 241  | 301 | 1.25 |
| 13 | Ashtabula | Youngstown | 86   | 90  | 1.05 |
| 14 | Cambridge | Cleveland  | 162  | 200 | 1.23 |
| 15 | Cambridge | Columbus   | 119  | 129 | 1.08 |
| 16 | Cambridge | Dayton     | 220  | 250 | 1.14 |
| 17 | Cambridge | Toledo     | 258  | 367 | 1.42 |
| 18 | Cambridge | Youngstown | 141  | 208 | 1.48 |
| 19 | Cleveland | Columbus   | 200  | 232 | 1.16 |
| 20 | Cleveland | Dayton     | 282  | 343 | 1.22 |
| 21 | Cleveland | Toledo     | 176  | 192 | 1.09 |
| 22 | Cleveland | Youngstown | 97.5 | 121 | 1.24 |
| 23 | Columbus  | Dayton     | 101  | 113 | 1.12 |
| 24 | Columbus  | Toledo     | 192  | 238 | 1.24 |
| 25 | Columbus  | Youngstown | 230  | 282 | 1.23 |
| 26 | Dayton    | Toledo     | 216  | 251 | 1.16 |
| 27 | Dayton    | Youngstown | 327  | 393 | 1.2  |
| 28 | Toledo    | Youngstown | 245  | 288 | 1.18 |
|    |           |            |      | Avg. Ratio= | 1.22 |

**Figure I:** Histogram of Ratio from Table II

(Attached Separately)

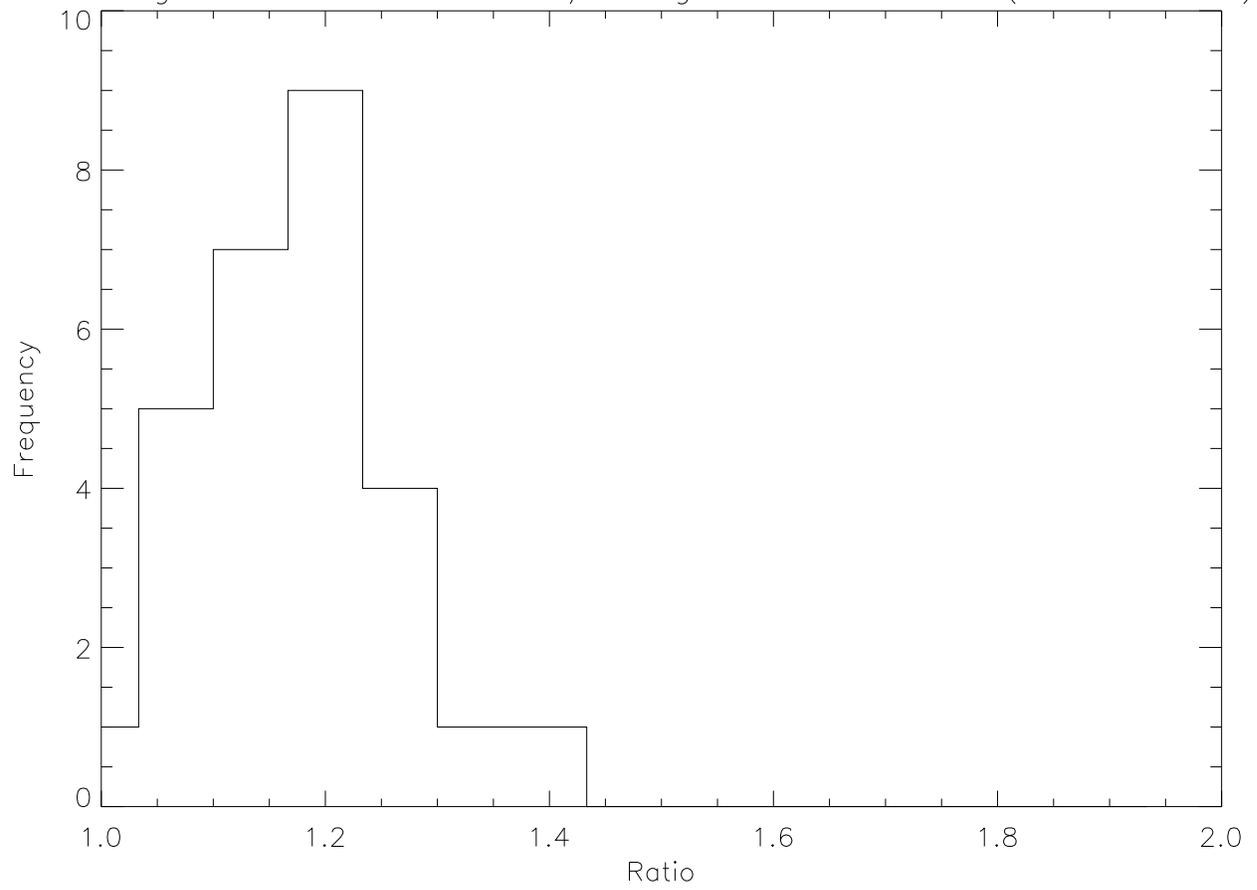